# Agentic Society：Merging skeleton from real world and texture from Large Language Model

Yuqi Bai[1]    Kun Sun[1*]    Huishi Yin[1]

1 Department of Computer and Information Engineering, Hebei Petroleum University of Technology, Chengde 067000, China.

* corresponding author email: heishuojinjsj@qq.com

## Abstract

Recent advancements in large language models (LLMs) and agent technologies offer promising solutions to the simulation of social science experiments, but the availability of data of real-world population required by many of them still poses as a major challenge. This paper explores a novel framework that leverages census data and LLMs to generate virtual populations, significantly reducing resource requirements and bypassing privacy compliance issues associated with real-world data, while keeping a statistical truthfulness. Drawing on real-world census data, our approach first generates a persona that reflects demographic characteristics of the population. We then employ LLMs to enrich these personas with intricate details, using techniques akin to those in image generative models but applied to textual data. Additionally, we propose a framework for the evaluation of the feasibility of our method with respect to capability of LLMs based on personality trait tests, specifically the Big Five model, which also enhances the depth and realism of the generated personas. Through preliminary experiments and analysis, we demonstrate that our method produces personas with variability essential for simulating diverse human behaviors in social science experiments. But the evaluation result shows that only weak sign of statistical truthfulness can be produced due to limited capability of current LLMs. Insights from our study also highlight the tension within LLMs between aligning with human values and reflecting real-world complexities. Thorough and rigorous test call for further research. Our codes are released at https://github.com/baiyuqi/agentic-society.git

## 1. Introduction

Social scientific experiments often incur substantial costs and extended durations due to the necessity of involving numerous human participants. However, advancements in large language models and related agent technologies are raising hope of offering promising solutions to mitigate these challenges, even though they cannot yet serve as immediate replacements[1-6]. The scale law and the emergence of abilities such as reasoning and planning enable large language models and agent technologies to exhibit characteristics of rationality[7-10]. Surprisingly, these models and agents can also be endowed with distinct personalities[11-14]. These cutting-edge advancements are raising hope or establishing a solid foundation for potential applications across a wide range of domains, including software development through agentic role-playing processes, non-player-character gaming, and the simulation of social science experiments, which is the primary focus of our study.

**Generation of a virtual population**

  The first task of most, if not all, significant social experiments is to sample from a population. This sampling process is crucial to the experiment's effectiveness and validity and is often the most resource-intensive component. Some emerging research involves using large language model agents to simulate social experiments by utilizing population data generated by real-world institutions[3]. But for most researchers or research teams, the accessibility of this real-world data poses a significant challenge due to limited resource and privacy regulations.

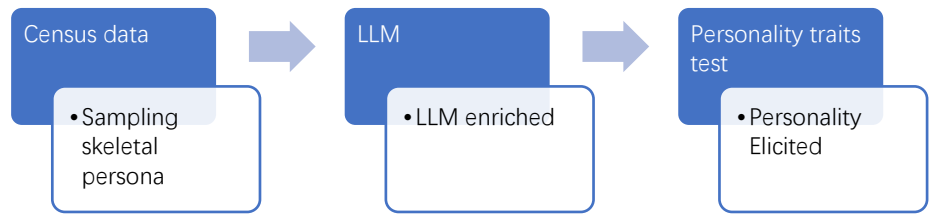

**Figure 1.** The framework of our method: 1)A feature vector(skeletal persona) is sampled from an engine based on census data 2) this skeletal persona is enriched with LLM, 3) personality traits test is applied on this persona and the result is used to further eliciting a much more deep persona

Most social science experiments focus on the results of statistical analyses of investigation data rather than the absolute accuracy, the ground truthfulness of individual data within the targeted population. In other words, social science experiments primarily require "statistical truthfulness".

The extensive corpus of data used to train large language models enables them to emulate human behaviors and preferences across various domains, including life style preferences, personality, even political attitudes. We define these collective aspects of an individual as a "persona.", and 1) the dimensions of this persona are huge 2) the interdependencies among these dimensions are even more complex. 3）those dimensions and interdependencies (conditional distribution) between them are implicitly incorporated in the large language model's text outputting probability, analogous to pixel dependencies in image and video generative models. Sampling from the generative models produce data with statistical truthfulness and sufficient variance.

To address the limitations of the aforementioned method we propose a novel approach that generates an agentic population with minimal resource requirements and no privacy compliance issues.

The capabilities embedded within large language models require careful elicitation, which is the essence of prompt engineering. More advanced techniques in this field continue to focus on effectively eliciting these capabilities. Again, just as image generative models works perfectly at enriching textural details of a sketch, intuitively we can use a skeletal persona to elicit details from large language model. In our research, we generate this skeleton by sampling from a statistical model based on real-world census data. This approach serves as a bridge between the real world and the large language model, and contributes significantly to the statistical truthfulness.

**Evaluation of the feasibility of our method with respect to LLMs**

The current body of research on the capabilities of large language models (LLMs) and their feasibility for simulating social science experiments is fraught with contradictions. Recent studies, in particular, have highlighted the limitations of LLMs in modeling human behavior, indicating that they are not yet adequately equipped for this task. We believe that LLMs are still in a relatively early and rapidly evolving stage of development. To assess our method of generating virtual populations, we propose an evaluation framework to determine whether existing and future LLMs are sufficiently advanced to produce virtual populations that exhibit a certain degree of "statistical truthfulness." Additionally, the ability of LLMs to simulate personality traits is one of the most extensively studied subfields in this domain, as personality is a fundamental aspect of human behavior. Therefore, our evaluation method is based on the simulation of personality traits.

Our contributions:
1) We propose a framework for generating a population to support large language model-enabled social science experiment simulations, significantly reducing resource requirements that are often inaccessible to many researchers.
2) We Implement a statistical model to sample persona from to make a population that is compliant with demographic properties of real-world population
3) We propose a method for using the Big Five personality traits test to elicit and enhance the personality profiles of the generated personas
4) We propose an statistical truthfulness evaluation framework and demonstrate the effectiveness and validity of large language models in enriching the details of personas but limited capability to capture the statistical truthfulness.
5) We reveal the tension within LLMs between aligning with human values and reflecting real-world complexities.

## 2. Related work

As we know it, there is still no research that is dedicated to the problem of the generation of a virtual agentic population to be used in agentic social scientific experiments, but some are highly relevant and should be viewed as lay the foundation of our research.

The ultimate foundation of every domain of LLM based agent is the emergence of such capabilities as reasoning and planning of the large language model, with various prompting techniques being collected to be named prompt engineering[7-10] . Some aspect of the broad research domain of LLM based agent are of interest to us, especially those concerning with the shaping of an agentic persona.[5, 12, 15-19]. Several studies have investigated methods of embedding personas into large language models by structuring separate and modular units within the model, rather than fine-tuning the entire model[13]. These approaches have shown superior effects compared to prompt-based methods. By utilizing modular units, researchers have achieved more efficient and effective persona integration, highlighting a promising direction for future advancements in large language model development.

Social science, distinct from natural science, focuses on human behavior and societal phenomena. Central to its methodology is individualism, which serves as a primary epistemological foundation. The validity of simulating social scientific experiments relies on the assumption that agent-based models, inspired by large language models (LLMs), accurately mimic human behavior. Our research is particularly interested in studies exploring the effectiveness and reliability of this imitation.[1, 11-14]

[11] provide rigor test and analysis of the construct validity of classic personality test on agentic personality, their work is the precondition of our research without which everything is illusion. The construct validity is itself a rigorous proof of the effectiveness of large language model in imitation of human personality, or the rigorous proof of the exitance of agentic personality in a Turing-test sense. [14] takes the personality traits as gene of LLM agent in determining its behavioral traits in game theory and its evolutionary dynamic, which enhanced this precondition.

The application of LLM agent in simulation of social science emerges as new trend in research[3, 4, 6], which is the domain to which our research makes contribution. This kind of research needs an agentic population to make the simulation, [3] takes real data (user profile and social networking data) from social media platforms, which is not accessible to most researchers. Also, privacy regulation compliance poses as an obstacle, [3] alleviates this issue by preprocessing the data by masking.

From prompt techniques to more advanced and complicate arrangements to stimulate large language models, we intuitively cognize a consistent effort to elicit knowledge and capabilities deeply hidden in the large language model[20, 21]. The agent role-playing interaction itself is a story of eliciting

In domain of social scientific experiment simulation, we as researchers might take vastly different perspectives with other LLM agent applications scenarios. [16] finds an alignment problem of large language model: it aligns with human value in question answering but demonstrate bias towards disabled (when playing the role of disabled, the large language model "thinks" itself void of certain mathematical capabilities just because itself being a disabled). This really is a bias in application scenarios such as character playing of agent in game (non-character playing) [22]or chatting with users, but in social science simulation, this bias might be an honest reflection of bias existed in reality. What if the target of the investigation is this bias in our society?

**Research on the limitation or unreliability of LLMs in simulation of human behaviors**

Recent works indicate that current large language models have significant limitations in simulating human behavior[23, 24]. These studies suggest that applications of large language models in the field of social science simulation experiments should be approached with necessary caution, and their implementation depends on further improvements in the models' capabilities.

3. Our method

**Real world sampling engine**

Our approach utilizes real-world census data [25] to develop a sampling engine that generates personas. These personas feature attributes such as age, gender, education, occupation, country, locality, annual income, and capital gains and losses. This is a skeleton of a persona, while each persona is virtual as an individual, its statistical properties reflect real-world distributions, because the joint distribution of the random feature vector is consistent with the real-world distribution. It is a population with similar demographic characteristics of real-world population. making it statistically representative of the actual population.

**From skeletal persona to "textured" rich persona**

By using the skeletal persona created above as part of prompt for a large language model, we can enrich it with intricate and rich details, as illustrated in figure 1 and figure 2. This generative process leverages the language model's ability to predict successive tokens, capturing a vast array of human characteristics beyond the capabilities of traditional statistical models.

**Potential of Generating Personas Without Sampling**

A key question arises here: can a large language model generate a population of personas with real-world demographic distributions without using a statistically sampled skeleton? While we have not conducted this experiment due to computational constraints, it is unlikely to succeed. The reasons are: 1) large language models are not yet powerful enough to achieve this independently, and 2) the intricate information required is deeply embedded and needs to be elicited.

An analogy with image generative models illustrates this point: generating an image of an animal randomly results in realistic textures but an alien overall appearance. However, using a skeletal prompt ensures the desired outcome.

**Three stage**

The framework can be view as a three-stage process: we have ground truth in census data, statistical truthfulness in skeletal persona, expanded statistical truthfulness in large language model enriched persona.

**Personality test and further elicitation of persona**

The human persona is multifaceted. Our method aims to generate a population for simulations in social science experiments across various disciplines, including psychology, behavioral economics, and sociology, as well as for commercial platforms in marketing research. The behavioral traits required for these simulations are diverse and versatile. But personality traits are fundamental and viewed as gene of behavioral traits[14], thus much more rigorously studied[11]. The outcomes of most social scientific experiments are significantly influenced by the personalities of the respondents. Therefore, we have a compelling rationale for selecting personality tests as a foundational element in enhancing our method. The results of these personality trait tests are utilized to validate our approach.

Another important functionality is to further refine and enrich the LLM-generated persona. After completing the personality test, we combine the resulting personality traits with the previously generated persona to construct a prompt. This process elicits a deeper and more enriched profile, which can then be used as a 'gene' for downstream tasks.

**Statistical truthfulness evaluation of our method with respect to capabilities of LLMs**

Our intuition is that if an LLM can adequately capture a prominent and sufficiently structured statistical characteristic present in real-world data—such as the distribution patterns of personality traits across demographic dimensions—then this capability would indicate that the LLM possesses the potential to support statistical truthfulness. We take this intuition as a hypothesis and base our evaluation method on it.

Personality traits are latent variables that play a significant role in human behavior, and research on LLMs' ability to simulate human behavior has most thoroughly explored the domain of personality traits. Therefore, we have chosen to base our evaluation framework on personality testing. As previously mentioned, personality test results are also used to enrich character profiles, thus serving a dual purpose in our method.

Previous research on LLMs' ability to simulate human behavior, particularly in modeling human personality traits, has largely focused on construct validity, often comparing LLM outputs with real-world data at the individual level. In contrast, our approach proposes a method for generating virtual populations, with effectiveness and feasibility grounded in statistical rather than individual-level truth. Therefore, our evaluation method is structured as follows:
   a) We begin with a real-world population for which census data and personality test data are available.
   b) We implement our virtual population generation method to create a corresponding virtual population.
   c) We then conduct personality testing on this virtual population.
   d) Finally, we perform statistical analyses on the personality test data from both the real and virtual populations, comparing their statistical similarities. The degree of similarity is regarded as a measure of statistical truthfulness.

When selecting statistical analyses for comparison of personality test data, the overall mean of a particular personality trait, while informative, may lack sufficient robustness. A more rigorous measure is the variation of this personality trait along a certain dimension. The choice of this dimension is crucial, as it should significantly influence personality traits. For example, age is a well-documented factor in personality development, whereas occupation has a much weaker correlation. Thus, statistically analyzing and comparing the curve of personality trait variation along the age axis represents a sound design choice. In the study of personality traits, age has indeed emerged as a particularly relevant dimension, allowing us to draw upon related research to obtain the age-related variation of personality traits in the real population for our evaluation.

**Relaxed Statistical truthfulness evaluation**

One of the greatest challenges in this evaluation framework is obtaining census data and personality test data from the same real-world population, a task that is often difficult, if not impossible. To overcome this challenge, we seek a type of statistic that demonstrates stability across different regions, countries, and time periods. If such a stable and consistent statistic can be identified, it can serve as a reference for our evaluation method, thereby mitigating this difficulty.

Personality traits' variation with age appears to be a promising candidate for this purpose. Fortunately, this idea and intuition are supported by previous psychological research, which indicates that the curve representing personality traits' variation along the age dimension does not significantly differ across regions, countries, or time periods. This stability is not found in other dimensions, such as occupation. By selecting the age dimension for our evaluation framework, we ensure that even under relaxed data conditions, the evaluation results still hold sufficient reference value. This leads us to a modified evaluation framework:
   a) We begin with census data from a real-world population and personality test data from another population. It is not required that these two datasets come from the same real-world population.
   b) We implement our virtual population generation method to create a virtual population.
   c) We then conduct personality testing on this virtual population.
   d) Finally, we perform statistical analyses on the personality test data from both the real and virtual populations, focusing on the age dimension. The degree of statistical similarity is regarded as a measure of statistical truthfulness.

The purpose of our evaluation method is to assess the inherent capability of a specific LLM to support statistical truthfulness. If this particular LLM can generate a virtual population and accurately simulate the personality traits of all individuals within that population, the consistency of these simulated traits with real-world statistics strongly suggests that the underlying LLM is capable of supporting social science experiment simulations that require statistical truthfulness as a fundamental criterion. It is important to note that this inherent capability of an LLM does not, by itself, constitute a sufficient condition for the success of a simulation application; the feasibility and effectiveness of specific applications still depend on various other factors. The evaluation framework does not aim to cover all aspects of LLMs' ability to simulate human behavior but rather seeks to provide a sufficiently effective indicator of this capability.

## 4. Experiment setup

### Skeletal persona Sampling engine based on census data

We utilize the adult income dataset in developing the skeletal persona sampling engine. Although this dataset is outdated and may not accurately reflect current demographics, it is widely used in research and numerous Jupyter notebooks. The skeletal persona sampling engine can generate samples either randomly or conditionally, with the latter being particularly useful for experiments targeting specific population groups.

### Large language model

The large language model employed in our research is the glm-4 model, accessed via a subscription-based open-platform API. By integrating this API with the LangChain framework, we ensure a seamless transition to alternative large language models, thereby enhancing flexibility and adaptability.

Default temperature value(0.7) is used in all requests.

### Personality trait test

The Big Five personality traits theory is a well-established framework in personality traits research, and is widely adopted in personality-related researches of large language models[11-14],. Our study employs the Big Five theory, specifically utilizing the IPIP-NEO-120 question inventory developed by Professor **Dr. John A. Johnson**, the shorter version of the IPIP-NEO-300 model created by **Lewis R. Goldberg.** The personality-related code in our research has been adapted and modified based on the work of [26]

To minimize resource expenditure, we initially attempted to submit all 120 questions in a single request to the GLM-4 model. However, this approach resulted in only partial responses, with approximately 50 questions answered. A similar issue occurred with ChatGPT, where only 101 questions were answered. Consequently, we divided the 120 questions into six groups of 20. This method effectively reduced costs while ensuring well-formed responses from the model. The responses from the six groups were then parsed and merged for processing by the IPIP-NEO engine to generate personality traits.

Finally, the personality traits derived from the IPIP-NEO engine were integrated with the pre-existing persona. This combined data was then used as a prompt to enhance and refine the persona.

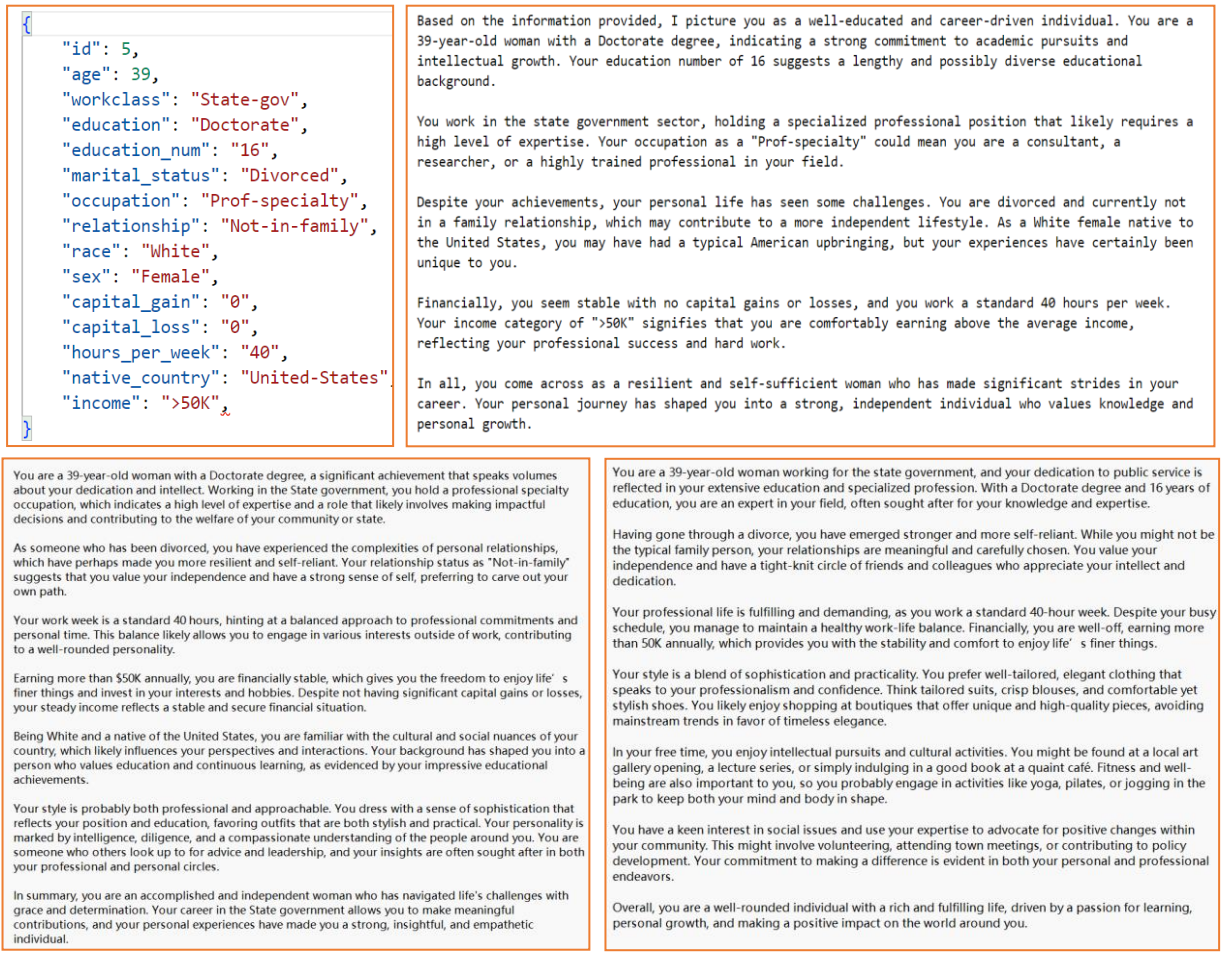

**Figure 2.** Skeletal features generated by sampling from engine based on census data (upper left) and LLM enriched persona based on it. Below is a comparison of two responses of the same skeletal persona feature

**Number of persona samples**

In our preliminary experiment, we utilized approximately 60 persona samples. Despite the modest scale of this setup, our primary objective at this stage was to observe the microscopic properties of our method and to conduct a fundamental statistical analysis.

**Statistical truthfulness evaluation**

As is discussed previously this evaluation can only be a partial evaluation. At the current stage of our research, we think some preliminary work of illumination, though insufficient, are valuable. We utilize data from [27] as the real-world dataset for comparison with our simulated personality traits data. Ideally, a rigorous analysis would require this dataset to be identical to the real-world population data that supports the persona skeletal sampling engine[25]. However, due to limited resources, this requirement cannot be fully met. Despite this discrepancy, we consider it acceptable, as our analysis aims to provide only partial evidence of the method's validity.

1. Results

**Skeleton and LLM enriched persona**

One sample of the generated persona skeleton and the corresponding LLM-enriched one is illustrated in figure 2. Scrutinizing through some of them makes it clear that details are greatly enriched, many of this enrichment are quite impressive. This enrichment is versatile and hard for enumeration, like life pattern, personality, dress and hair style. Unrelated facts in the skeletal persona are put into causal relationship, like the single life of a divorced man spending more time on his work leads to a high workhour.

Another thing worth noting and important is the fact that the same skeleton produces substantially different text, though it is predictable. Through experiment our insight is that this difference is not just superficial wording of the same matter, it is the language model

> Here's the breakdown of the choices:
>
> - **Question 1**: "Am attached to conventional ways." Assuming this professional follows a structured, perhaps traditional career path, a 3 (Neither Accurate Nor Inaccurate) is chosen.
> - **Question 2**: "Think highly of myself." Given the background of success and a strong work ethic, a 4 (Moderately Accurate) seems appropriate.
> - **Question 3**: "Carry out my plans." As a responsible professional, a 5 (Very Accurate) is likely.
> - **Question 4**: "Dislike myself." This is likely 1 (Very Inaccurate) given the positive self-image implicit in the background.
> - **Question 5**: "Have a high opinion of myself." Similar to Question 2, a 4 (Moderately Accurate) is chosen.
> - **Question 6**: "Rarely notice my emotional reactions." Given the nature of government work, being aware of emotions is important, so a 2 (Moderately Inaccurate).
> - **Question 7**: "Get back at others." This is not indicative of a responsible randomly pays its attention to some aspect of the persona. To the fact that she is devoiced, the second generated persona depicted her " tight-knit circle of friends".

**Figure 3.** The LLM provides rationale behind its choices of IPIP-NEO 120 questions

**Personality test result and its analysis**

We sliced the IPIP-NEO 120 question inventory into six group to produce a quiz sheet as part of the prompt. The response from the large language model demonstrates its rationale behind its choices, illustrated in figure 3

At times, the language model responds with a randomly selected answer, suggesting that the prompting is insufficient. To address this, we inserted strong directive words to compel the model to provide the rationale behind its choices, which proved effective. The large language model is forced to "think" by text.

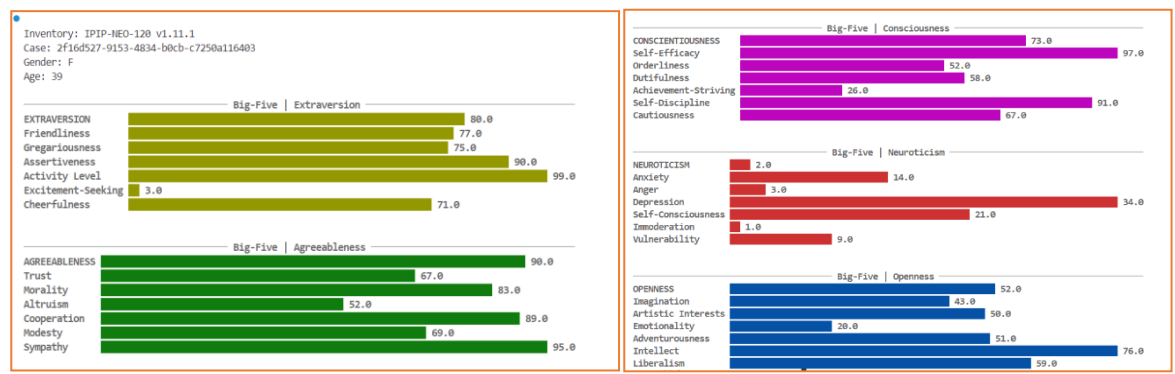

**Figure 4.** The Big Five personality traits (the same persona as in figure 1.)

The big five personality traits each with its sub traits of the sampled persona provided in figure 2 are illustrated in figure 4.

> You are an engaging and dynamic individual, brimming with energy and a zest for life. With a high score in extraversion, you thrive in social settings, where your charm and ease with people allow you to effortlessly make connections and friends. Your enthusiasm for adventure and new experiences is contagious, and you often find yourself at the heart of any gathering, ready to introduce others to the joy of the moment.
> Your agreeableness is a testament to your compassion and cooperative spirit. You are the sort of person who always has time to lend a hand, and your honest and caring nature endears you to those around you. Your ability to see the best in others fosters an environment of trust and harmony, making you a valuable mediator and a beloved friend.
> Conscientiousness is one of your defining traits; you are meticulous in your approach to life, valuing structure and order. Your goal-driven nature propels you to success, as you persistently work towards achieving your objectives with a determination that is both admirable and effective. Colleagues and peers regard you as highly reliable and fair, qualities that have surely contributed to your professional achievements.
> Neuroticism is not a dominant aspect of your personality; you maintain a calm and level-headed demeanor even in the face of stress. Your low score in this trait suggests a resilience and an ability to keep a positive outlook, which serves you well in navigating life's challenges.
> In terms of openness, you strike a balance between curiosity and practicality. While you enjoy exploring new ideas and approaches, your average score indicates a grounded nature that keeps you rooted in reality. This balance allows for creativity while ensuring that you maintain a focus on tangible outcomes.
> As a well-educated professional, you have dedicated yourself to a life of learning and personal betterment. Your journey has not been without its difficulties, but your independence and strength have seen you through. You are a divorced woman who has embraced the challenges of singlehood, using them to grow into a self-sufficient and successful individual.
> Financially stable and intellectually stimulated, you are a beacon of confidence and capability. Your persona is one of an accomplished, lively, and deeply caring individual, ready to take on the world with a smile and an unwavering spirit.

**Figure 6.** Personality elicited persona(same person as in figure 2 and figure 4)

**Personality elicited persona**

Personality traits test is a psychological experiment based on the persona, but can also be viewed as an eliciting process of the persona itself, so we use the big five IPIP-NEO 120 evaluation result(in Json format) combined with the basic persona to prompt the large language model to generate a response that greatly enriched existent persona with a much comprehensive personality. We also prefixed the prompt with a piece of text defining the meaning of the big five personality traits to enhance the large language model understanding. The result can be viewed in figure 6

In a large portion of social science experiments personality traits are basic and critical elements, this fact might make the above persona eliciting an essential step in the generation of persona.

**Statistical truthfulness evaluation of glm-4 model**

The change curve of personality traits along the age dimension in the real population is derived from [27]. However, due to the unavailability of the original data, we utilized the mean values of the Big Five personality traits across age groups as provided in the appendices of the referenced paper. Corresponding to this accessible statistical analysis of real population data, we conducted a parallel statistical analysis of the personality simulation results generated by the GLM-4 language model. We then performed a similarity analysis between them, using this comparison as a measure of statistical truthfulness when applying our method to GLM-4.

At the current stage, our analysis combines similarity calculations with direct visual inspection to gain insights.

**Measure of the Statistical truthfulness evaluation**

1. Due to the unavailability of the original personality test data from the real population as described in [27], we relied on the mean values of personality traits segmented by age groups, as provided in the paper, for similarity calculations. Consequently, the

GLM-4 personality test data were aggregated to reflect the mean values across the corresponding age groups.
2. For the similarity calculation, we employed the Euclidean distance method, which offers a straightforward approach.

The results of the calculations are provided in table 1 to 4.

| age_range | extraversion | agreeableness | conscientiousness | neuroticism | openness |
|---|---|---|---|---|---|
| 16_19 | 53.01 | 48.61 | 42.76 | 50.47 | 50.45 |
| 20_29 | 51.58 | 50 | 47.88 | 50.1 | 51.08 |
| 30_39 | 49.7 | 50.43 | 50.35 | 49.92 | 49.79 |
| 40_49 | 48.54 | 50.91 | 50.82 | 49.39 | 48.64 |
| 50_59 | 47.47 | 51.32 | 50.8 | 48.99 | 48.06 |
| 60_69 | 46.98 | 50.98 | 49.24 | 47.87 | 46.28 |
| 70_79 | 45.56 | 51.43 | 47.2 | 46.25 | 44.27 |
| 80_85 | 45.41 | 51.44 | 46.77 | 46.52 | 42.47 |

Table 1. BHPS big five along age group

| age_range | extraversion | agreeableness | conscientiousness | neuroticism | openness |
|---|---|---|---|---|---|
| 16_19 | 51.17 | 49.64 | 41.49 | 48.8 | 51.75 |
| 20_29 | 50.94 | 49.65 | 47.15 | 49.99 | 51.46 |
| 30_39 | 50.12 | 49.79 | 50.22 | 50.04 | 50.23 |
| 40_49 | 49.84 | 50.31 | 51.22 | 50.36 | 50.15 |
| 50_59 | 49.08 | 50.21 | 51.16 | 51.1 | 50.43 |
| 60_69 | 48.27 | 50.56 | 50.23 | 51.51 | 49.43 |
| 70_79 | 47.54 | 52.46 | 50.46 | 51.38 | 47.66 |
| 80_84 | 47.57 | 54.16 | 49.84 | 50.74 | 45.56 |

Table 2. GSOEP big five along age group

| age_range | extraversion | agreeableness | conscientiousness | neuroticism | openness |
|---|---|---|---|---|---|
| 16_19 | 49.06457 | 92.44805 | 87.37081 | 6.217751 | 45.05474 |
| 20_29 | 56.08587 | 93.00299 | 88.20594 | 10.60342 | 27.7755 |
| 30_39 | 47.84194 | 92.7612 | 92.70027 | 10.83149 | 21.75161 |
| 40_49 | 51.05459 | 91.95088 | 94.14761 | 13.51718 | 20.60066 |
| 50_59 | 58.50002 | 89.58505 | 91.32356 | 15.70212 | 23.95549 |
| 60_69 | 44.33395 | 93.61068 | 89.3246 | 19.03558 | 14.09337 |

Table3. glm-4 simulated Big Five along age dimension

| data-sourse | extraversion | agreeableness | conscientiousness | neuroticism | openness |
|---|---|---|---|---|---|
| bhps-gsoep | 1.29 | 0.75 | 0.76 | 1.89 | 1.82 |
| glm4-bhps | 5.39 | 41.89 | 41.9 | 37.13 | 25.04 |
| glm4-gsoep | 4.86 | 42.22 | 41.99 | 37.79 | 26.58 |

Table 4. The Euclidean distance between the GLM-4 model and the real population data for the Big Five personality traits across the age dimension indicates that the simulated data closely approximates the real data only in the trait of extraversion.

BHPS and GSOEP are personality test datasets from different countries, with detailed information provided in [27]. These distances were calculated using data extracted from the appendix of [27], and the extracted data are also provided in the table 1 and table 2 of this paper. The first row of Table 4 presents the Euclidean distances between the Big Five personality traits' age-related change curves in these two real-world populations. The subsequent rows display the Euclidean distances between the change curves generated by GLM-4 and those of the two real-world datasets.

The experimental results indicate that, except for extraversion, where the simulated data closely aligns with the real data, the other four personality traits show significant deviations between the simulated and real data.

**Conspicuous bias of LLM in personality trait test**

To gain intuitive insights, we visualized the change curve of the simulated personality data across the age dimension and compared it with the corresponding changes in the real data (with the real data trends provided in the appendix C).

The Big Five personality traits across different age groups of simulation is illustrated in Figure 5. Several key insights have been identified: 1) The curve for conscientiousness demonstrates

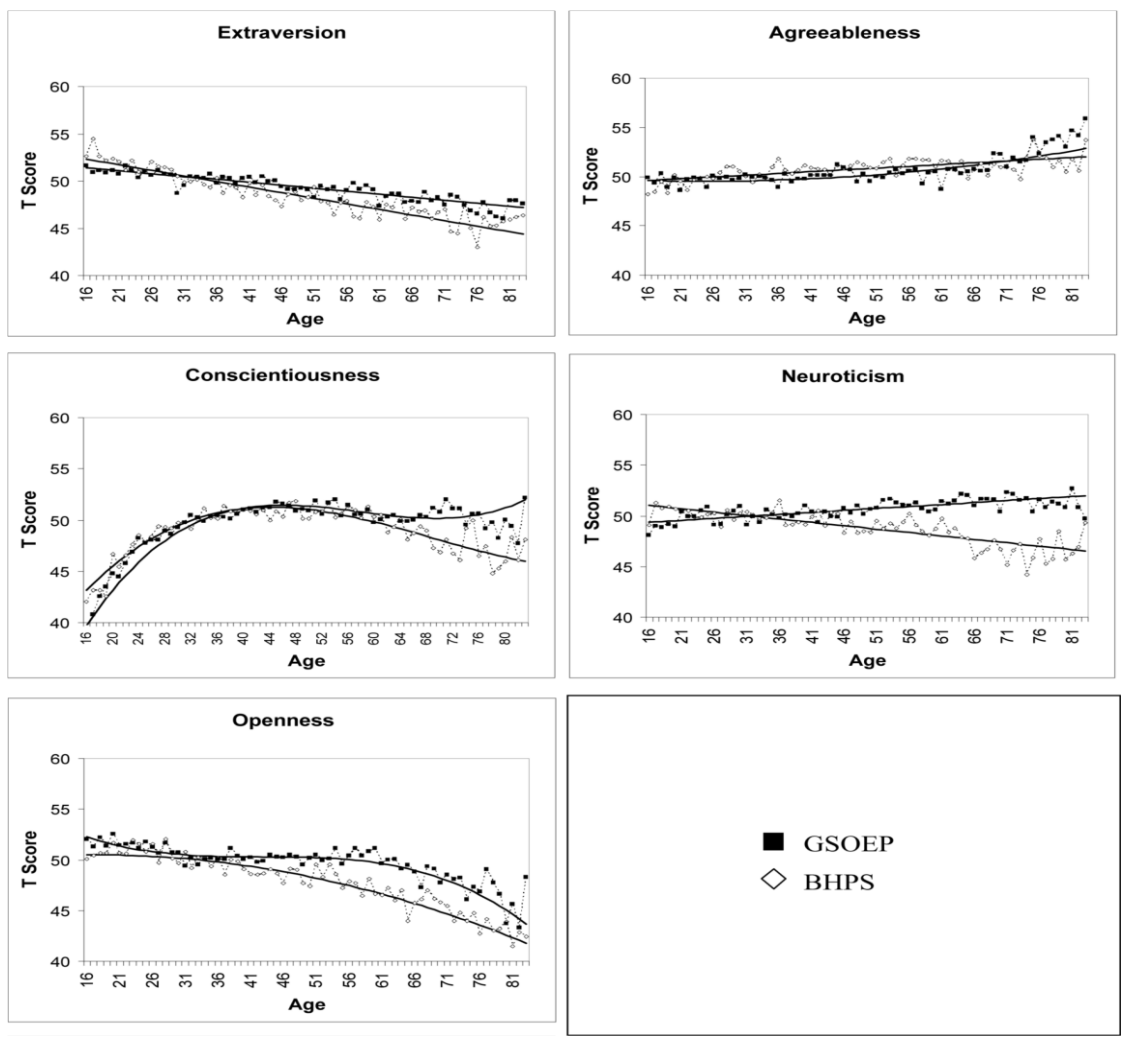

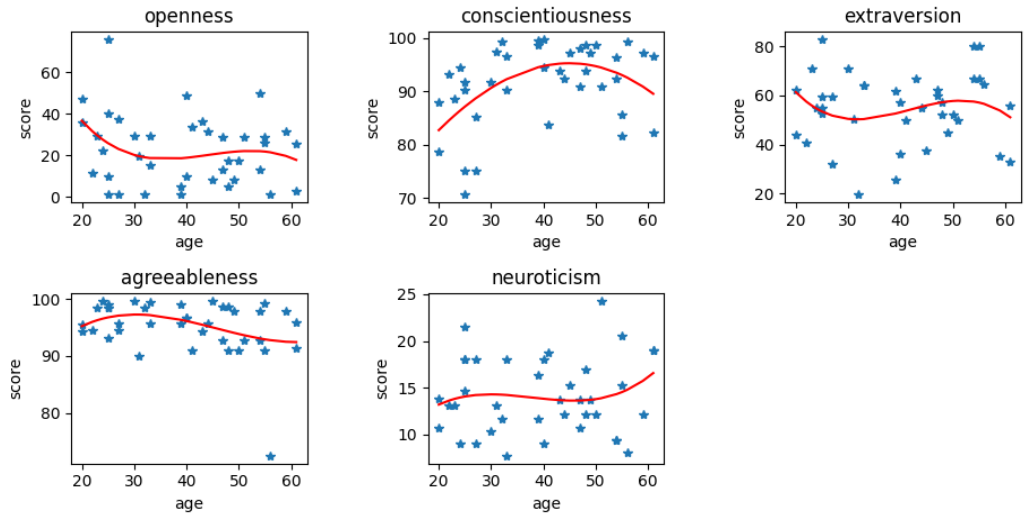

**Figure 5.** The development of the five personality traits along the age dimension. Top is that of real population, bottom is data simulated by glm-4

a reasonable pattern, showing an increase from the 20s to the 50s, which aligns with the development of responsibility over time. 2) Both openness and extraversion
exhibit similar patterns, consistent with intuitive expectations and their observed similarity. 3) Notably, the scores for conscientiousness and agreeableness are significantly higher than existing statistical results for the general population, while the score for neuroticism is conspicuously lower. This discrepancy warrants further investigation.

Further observations reveal that the traits with higher scores, conscientiousness and agreeableness, are associated with "positive qualities," while the trait with a lower score, neuroticism, is considered a "negative quality." This suggests that the large language model is optimized to exhibit well-behaved traits, is doing its best to be well-behaved. Such behavior likely results from efforts to align the model with human values and responsible AI practices. The model's responsibility to reflect positive human values appears to be a repression for it to play its assigned role in the personality trats testing.

This reveals a problem fundamental in theory and practice of human value alignment or responsible AI. In many scenarios it is necessary and a merit that large language model speaks pleasantly, but in much more scenario we need it being truthful to reality, not political correctly. Strictly defining these concepts reveals a crucial distinction: a language model behaving in an idealized manner does not necessarily imply that it perceives everyone as ideal. However, even humans frequently lose sight of this logical distinction. This is quite a fundamental intrinsic tension of large language model, what [16] investigated, but from a perspective on the contrary.

## 5. Conclusion

We propose a framework for generating a population to support simulations of social science experiments using LLM agents, aiming to significantly reduce resource requirements that are typically inaccessible to many researchers. We then propose a statistical truthfulness evaluation framework to evaluate the capability of LLMs to capture the statistical truthfulness. While a thorough and rigorous experiment on various large language models has not yet been conducted in current stage, preliminary experiments on glm-4 and analyses indicate the

following findings:
1. At a microscale, the generated personas exhibit rich diversity across various aspects of human life, characterized by sufficient randomness crucial for such experiments.
2. Evaluation result shows that only weak sign of statistical truthfulness can be produced due to limited capability of current LLMs.
3. Importantly, insights derived from the statistical results highlight a fundamental aspect of large language models: the inherent tension between adhering to correct human values and accurately reflecting human realities.

**Data availability**

The datasets generated during and/or analysed during the current study are available in the github repository, https://github.com/baiyuqi/agentic-society.git

**References**

1. Aher, G.V., R.I. Arriaga, and A.T. Kalai. *Using large language models to simulate multiple humans and replicate human subject studies*. in *International Conference on Machine Learning*. 2023. PMLR.
2. Chen, H., et al., *RoleInteract: Evaluating the Social Interaction of Role-Playing Agents.* arXiv preprint arXiv:2403.13679, 2024.
3. Gao, C., et al., *S $^ 3$: Social-network Simulation System with Large Language Model-Empowered Agents.* arXiv preprint arXiv:2307.14984, 2023.
4. Horton, J.J., *Large language models as simulated economic agents: What can we learn from homo silicus?* 2023, National Bureau of Economic Research.
5. Hu, T. and N. Collier, *Quantifying the Persona Effect in LLM Simulations.* arXiv preprint arXiv:2402.10811, 2024.
6. Li, N., et al., *Large language model-empowered agents for simulating macroeconomic activities.* arXiv preprint arXiv:2310.10436, 2023.
7. Wang, L., et al., *Plan-and-solve prompting: Improving zero-shot chain-of-thought reasoning by large language models.* arXiv preprint arXiv:2305.04091, 2023.
8. Wei, J., et al., *Emergent abilities of large language models.* arXiv preprint arXiv:2206.07682, 2022.
9. Xu, B., et al., *Rewoo: Decoupling reasoning from observations for efficient augmented language models.* arXiv preprint arXiv:2305.18323, 2023.
10. Yao, S., et al., *Tree of thoughts: Deliberate problem solving with large language models.* Advances in Neural Information Processing Systems, 2024. **36**.
11. Safdari, M., et al., *Personality traits in large language models.* arXiv preprint arXiv:2307.00184, 2023.
12. He, Z. and C. Zhang, *AFSPP: Agent Framework for Shaping Preference and Personality with Large Language Models.* arXiv preprint arXiv:2401.02870, 2024.
13. Li, T., X. Zheng, and X. Huang, *Tailoring personality traits in large language models via unsupervisedly-built personalized lexicons.* arXiv preprint arXiv:2310.16582, 2023.
14. Suzuki, R. and T. Arita, *An evolutionary model of personality traits related to cooperative behavior using a large language model.* Scientific Reports, 2024. **14**(1): p. 5989.

15. Cheng, M., E. Durmus, and D. Jurafsky, *Marked personas: Using natural language prompts to measure stereotypes in language models.* arXiv preprint arXiv:2305.18189, 2023.
16. Gupta, S., et al., *Bias runs deep: Implicit reasoning biases in persona-assigned llms.* arXiv preprint arXiv:2311.04892, 2023.
17. Han, X., et al., *PersonaPKT: Building personalized dialogue agents via parameter-efficient knowledge transfer.* arXiv preprint arXiv:2306.08126, 2023.
18. Li, J., et al., *A persona-based neural conversation model.* arXiv preprint arXiv:1603.06155, 2016.
19. Li, J., et al., *On the steerability of large language models toward data-driven personas.* arXiv preprint arXiv:2311.04978, 2023.
20. Burns, C., et al., *Weak-to-strong generalization: Eliciting strong capabilities with weak supervision.* arXiv preprint arXiv:2312.09390, 2023.
21. Lu, K., et al., *Large Language Models are Superpositions of All Characters: Attaining Arbitrary Role-play via Self-Alignment.* arXiv preprint arXiv:2401.12474, 2024.
22. Hu, S., et al., *A survey on large language model-based game agents.* arXiv preprint arXiv:2404.02039, 2024.
23. Petrov, N.B., G. Serapio-García, and J. Rentfrow, *Limited Ability of LLMs to Simulate Human Psychological Behaviours: a Psychometric Analysis.* arXiv preprint arXiv:2405.07248, 2024.
24. Shu, B., et al., *You don't need a personality test to know these models are unreliable: Assessing the Reliability of Large Language Models on Psychometric Instruments.* arXiv preprint arXiv:2311.09718, 2023.
25. Becker, R.K.a.B., *Census Income.* 1994.
26. Corbari, E. *Five Factor E*. Available from: https://github.com/NeuroQuestAi/five-factor-e.
27. Donnellan, M.B. and R.E. Lucas, *Age differences in the Big Five across the life span: evidence from two national samples.* Psychology and aging, 2008. **23**(3): p. 558.

## Author contributions

Yuqi Bai: conceptualization, methodology, data processing, writing—original draf. Kun Sun: Writing—review & editing. Huishi Yin: data processing. All authors have read and agreed to the submitted version of the manuscript.

## Competing interests

The authors declare no competing interests.